\documentclass[preprint,prl,showpacs,preprintnumbers,amsmath,amssymb]{revtex4}
\usepackage{amsmath}
\usepackage{graphics}
\usepackage{graphicx}
\usepackage{amsfonts}
\usepackage{amssymb}

\def\registered{{\ooalign
    {\hfil\raise.07ex\hbox{{\scriptsize\textup{R}}}\hfil\crcr\mathhexbox20D}}}
\begin{document}

\preprint{HEP/123-qed}
\title{Soret driven convection in a colloidal solution heated from above at very
large solutal Rayleigh number.}
\author{Roberto Cerbino, Alberto Vailati, and Marzio Giglio}
\affiliation{Istituto Nazionale per la Fisica della Materia (INFM) and Department of
Physics, University of Milan, Via Celoria 16, 20133 Milano, Italy.}

\begin{abstract}
Convection in a colloidal suspension with a large negative separation ratio $\psi$ is studied experimentally by heating from above. Shadowgraph observation at very large solutal Rayleigh numbers $\tilde{R}$ \ are reported as a function of time. Fast relaxation oscillations are reported for the root mean square value of the shadowgraph intensity. While pure fluids exhibit a transition to turbulent convection for Rayleigh number $R\simeq10^{6}$, stable spoke-pattern planform with up and down columnar flows are observed up to $\tilde{R}\simeq1.9\times10^{9}$. It is suggested that the surprising stability of the planform against turbulence is due to nonlinear focusing arising from the concentration dependence of the diffusion coefficient.

\end{abstract}
\volumeyear{year}
\volumenumber{number}
\issuenumber{number}
\eid{identifier}
\date[Date text]{date}
\received{date}
\revised{date}
\accepted{date}
\published{date}
\startpage{101}
\endpage{102}
\maketitle

Soret driven convective instabilities (SDI) may arise in a horizontal fluid mixture slab because the Soret induced concentration gradient is top heavy and destabilizing \cite{vs,nota}. It has been shown long ago that SDI in liquid mixtures with a small Lewis number $L=D/\kappa$ (D and $\kappa$ are the mass and the heat diffusion coefficient, respectively) are described by the same equations of the classical single component Rayleigh-B\'{e}nard convective instability (RBI) \cite{dg1}. Since heat diffuses much faster than mass, the temperature $T$ becomes an irrelevant variable, and the temperature gradient remains uniform even during convection, buoyancy forces arising only because of concentration changes. The equation describing SDI then become identical to those for RBI, the concentration diffusion equation replacing the temperature diffusion equation.

RBI has been extensively studied at large Rayleigh numbers because of its relevance in important phenomena, like the mantle convection \cite{bussemantle}, and because at even larger Rayleigh numbers it is a nice model for the transition to turbulence \cite{kadanoff}. Conversely, early studies on SDI have been conducted mostly close to the threshold \cite{hj,sparasci,gvprlsoret,moses} and only recently SDI has been explored for fairly large solutal Rayleigh numbers \cite{laporta}, but not quite high enough to explore the region corresponding to the transition to turbulent convection.

In the present work we have investigated a SDI at extremely high solutal Rayleigh numbers by using a very unorthodox sample, a dilute colloidal suspension with an unusually large negative separation ratio $\psi$ and small Lewis number $L$. Stresses from $\tilde{R}=1.88\times10^{6}$ up to $\tilde{R}=1.90\times10^{9}$ have been studied, heating being provided from above.
The results are very surprising. Even for the highest stress (a factor of 1000 above the equivalent for the transition to turbulence for RBI) stable spoke patterns with up and down columnar flow are observed . The exceptional stability against turbulent behavior will be tentatively justified on the basis of a non-linear focusing effect recently described in RBI simulations \cite{yuen2,yuen3}. We have also studied the time evolution of the convective modes amplitude by performing a quantitative analysis of shadowgraph images. We find that the convective flow sets in with a time constant that is orders of magnitude smaller than the mass diffusion time constant across the sample height $\tau_{m}=d^{2}/D$, where $d$ is the sample thickness. The convective mode amplitude shows damped oscillations with a period in very good agreement with some recent theoretical predictions \cite{shliomis}.

The most common configuration for RBI contemplates conducting boundaries and the threshold is $R=1708$ where \begin{equation} R=\frac{\alpha g\Delta Td^{3}}{\kappa\nu}\label{ray} \end{equation} is the Rayleigh number, $\alpha$ is the thermal expansion coefficient, $g$ the
acceleration of gravity, $\nu$ the kinematic viscosity and $\Delta T$ the imposed temperature difference over the sample height $d$. For SDI the boundaries are almost invariably impermeable (no solute particles can be released or absorbed at the boundaries) and the threshold condition is given by the solutal Rayleigh number $\tilde{R}=720$ where \begin{equation} \tilde{R}=\frac{\beta g\Delta c\ d^{3}}{D\nu}=\frac{-\psi}{L}R\label{solutal} \end{equation}
$\beta=\frac{1}{\rho}(\frac{\partial\rho}{\partial c})$ is the solutal expansion coefficient, $\psi$ is the separation ratio between density changes due to the Soret driven concentration changes and those due to the thermal expansion \cite{soret}. It should be remarked that for single component liquids, $R=720$ is the convective threshold for the non conducting boundaries, and this further stresses the strong formal similarity between RBI and SDI.

According to Eq.\ref{solutal} the SDI threshold temperature gradient is a factor $L/\psi$ lower than for RBI. So the strategy to be able to reach high stresses (still maintaining experimentally reasonable temperature difference between top and bottom plates) is to select a sample with $L/\psi$ as small as possible, and colloidal suspensions appear as good candidates, because $D$ is usually very small, and Soret effects are large \cite{morozov}. Also, a sample with a negative $\psi$ must be used, because then heating is required from above, and consequently there is no danger of inducing buoyancy forces due to the thermal expansion of the sample. Therefore, no RBI can be triggered no matter how large the applied temperature gradient is.

We have selected a dilute (c=4.1\% wt.) aqueous colloidal suspension of 11 nm radius silica spheres (LUDOX{\ooalign {\hfil\raise.07ex\hbox{{\scriptsize\textup{R}}} \hfil\crcr\mathhexbox20D} }\ TMA) characterized by $L=1.49\times10^{-4}$ and $\psi$=-4.06. When compared with ordinary liquid mixtures ($L\simeq10^{-2}$ and $\psi$ is of order $10^{-1}$\cite{kolodner}), these $L$ and $\psi$ values almost equally contribute to an overall reduction of the threshold by a factor of a thousand. Other properties of the sample (at $T=30$°C) are $\alpha=3.03\times10^{-4}$K$^{-1}$, $\kappa=1.48\times10^{-3}$cm$^{2}$/s, $\nu=8.18\times10^{-3}$cm$^{2}$/s, $D=2.2\times10^{-7}$cm$^{2}$/s, $\beta=0.69$, $k_{T}=-0.54$.

The samples have the shape of horizontal discs, 41.00 mm and 40.65 mm in diameter for d=0.98 mm and d=4.50 mm respectively. Top and bottom boundaries are two horizontal, 8 mm thick sapphire plates that are temperature controlled by a PID servo driving two annular Peltier elements. Temperature stability is better than a few mK over many hours. Horizontal temperature non uniformities across the inner surfaces at steady state have been estimated by means of numerical simulations to be less than 2\%. However, during the transient, radial gradient are unavoidably induced inside the sample. As a consequence, a slight radial anisotropy is observed in the convective patterns during the early stages of convection. The cell has been described in more detail elsewhere \cite{vg96}. Measurements with the thinner cell were taken for $\Delta$T=2.98 K, $\Delta$T=6.54 K, and $\Delta$T=13.22 K (corresponding respectively to $\tilde{R}=1.88\times10^{6}$, $\tilde{R}=4.12\times10^{6}$, $\tilde{R}=8.33\times10^{6})$. The largest stress value were obtained with the thicker cell at $\Delta$T=31.13 K ($\tilde{R}=1.90\times 10^{9}$).

All the measurements were performed by imposing at t=0 a sudden vertical temperature gradient across the cell and observing with a shadowgraph technique the time evolution of the intensity distribution of the transmitted beam on a plane 3.0 cm away from the sample. The probing beam is
directed vertically. A background image taken in the absence of a temperature gradient was subtracted to each frame to eliminate the effects of non-uniform illumination. In the shadowgraph images bright features are associated with regions where the colloid concentration is higher (colloid rich regions act as positive focal length lenses). Conversely, the dark features map colloid poor regions.

We report in Fig.1 a sequence of shadow images for the 0.98 mm sample and with $\Delta$T=13.22 K. The 90\% of the setpoint was reached in about 80 s and this time is included in the times stated below. After an induction time of 390 s where no perturbations were observed (see Fig.1a), the planform of the instability suddenly appeared across the entire sample (Fig.1b). This induction time is enormously smaller than the mass diffusion time scale $\tau_{m}$=4.4$\times$10$^{4}$ s. During the induction time, large concentration gradients are created in thin layers near the boundaries, the central region being at uniform concentration. Because the boundaries are impermeable to solute flow, the concentration gradient there attains the steady state value $\vec{\nabla}c=-\frac{k_{T}}{T}\vec{\nabla}T$ \ very rapidly, and the thickness of the layer grows like $\sqrt{Dt}$. Taking t=390 s, the layer thickness is of the order of 90 $\mu m$. The planform evolved rather rapidly, and eventually attained an almost stationary pattern after 7200 s and shown in Fig.1c. We have monitored the convective planform for 9 days corresponding to 18 mass diffusive time constants. We show the planform for this delay in Fig.1d. One can hardly notice changes
between Fig.1c and Fig.1d.

They both show the typical spoke pattern convective structure commonly observed for RBI at high stresses \cite{busse}, and also reported for SDI in water-alcohol mixtures \cite{laporta}. In spite of the fact that these structures are pretty ubiquitous in convective instabilities, to our knowledge no attempt has been made so far to find out what the three dimensional structure of the convective pattern actually is, and the shape of the shadow images is rather puzzling. We will attempt below to address this problem. Figures 1c and 1d are composed of irregular polygonal loops of thin bright and dark lines, staggered so that bright and dark lines tend to cross at right angles. Bright lines converge to bright nodes, and dark lines to dark nodes. Furthermore, bright nodes tend to fall near the centers of dark loops, and dark nodes near the centers of bright loops.

The very fact that bright and dark lines cross, strongly suggests that the corresponding perturbations are localized either close to the top or the bottom plates respectively, and the top-to-bottom shadow imaging superimposes the effects of the separate perturbations. The bright and dark lines correspond to mostly horizontal flows localized near the boundaries, and these flows eventually join at the nodes, where they promote convective flow across the entire layer.
Consequently, we believe that penetrative convection takes place in the form of columnar flow
localized at the nodes, down-flowing colloid rich regions corresponding to the bright nodes, and colloid poor up-flow occurring at the dark nodes.

All the above results are in strikingly good agreement with recent numerical simulations for high Prandtl RBI \cite{yuen}.

We present in Fig.2 a 3D image of the convective planform obtained with the simulation. A cold (blue) isotherm and a hot (red) isotherm are shown, and the upwelling and downwelling columnar convective flows are immediately apparent, as well as their spatial staggering. Also notice that close to the top and bottom boundaries, runners feeding the columnar flow are created, and they are in the form of spokes converging toward the bases of the
columnar structures. A detailed 2D mapping of the isotherms at two planes close to the boundaries (see Fig. 3 or Fig. 4 of Ref.\cite{yuen}, not shown here for brevity) do show that close to the boundaries the hot and cold runners tend to be arranged at right angles, very much as the bright and dark lines in the shadow images.

The shadow images in Fig.1 can be analyzed to obtain an estimate of the overall concentration modulations associated with the convecting mode. It can be shown that the root mean square of the intensity distribution is related to the amplitude of the spatial concentration modulation.

Images from the sequence in Fig.1 (referring to $\tilde{R}=8.33\times10^{6}$) and for the other two cases, $\tilde{R}=1.88\times10^{6}$, $\tilde{R}=4.12\times10^{6}$ have been analyzed. The result for the case $\tilde{R}=8.33\times10^{6}$ are shown in Fig. 3a, where the RMS is plotted as a function both of the reduced mass diffusion time $t/\tau_{m}$ and
as a function of the reduced thermal time $t/\tau_{h}$ (where $\tau_{h}=d^{2}/\kappa$). The curve shows an induction time (corresponding to the uniform gray image Fig.1a), then a peak and pronounced relaxation oscillations, and finally the approach of the steady state. The induction time is quite comparable with the oscillation period, and both are orders of magnitude shorter than the diffusive time $\tau_{m}$, and quite long compared with the thermal time constant $\tau_{h}$.

Recent theoretical work \cite{shliomis} predicts that strongly localized gradients as these do create a convective flow that eventually destroys the gradient. The gradient then builds back, and the process is repeated. The theory provides quantitative estimates for the oscillation period. We report in Table \ref{tabella} the experimental values for the oscillation period together with the theoretical estimates.
\begin{table}[tbp] \centering
\begin{tabular}
[c]{|l|l|l|}\hline
$\tilde{R}$ $(\times10^{-6})$ & $t_{obs}(s)$ & $t_{teo}(s)$\\\hline
$1.88$ & $550$ & $511$\\\hline
$4.12$ & $386$ & $374$\\\hline
$8.33$ & $299$ & $283$\\\hline
\end{tabular}
\caption{Oscillation period for different $\tilde{R}$ for d=0.98mm. $t_{teo}$ is calculated from Ref.\cite{shliomis} \label{tabella}
}
\end{table}
The agreement is very good, in spite of the fact that our experimental results show no trace of a self regenerative process as described in Ref.\cite{shliomis}, and show instead that a stable concentration mode is maintained after the relaxation oscillations.

Also shown in Fig.3b is the time evolution for the only run with the sample with thickness d= 4.50 and at very high stresses, $\tilde{R}=1.90\times10^{9}$ (Fig.3b). One can notice that the induction time has got even faster, three orders of magnitude smaller than the mass diffusive time constant, and almost comparable with the thermal time constant. No oscillations are observed, and while this may be purely accidental, the theoretical calculations of Ref.\cite{shliomis} fail to produce real roots for the rate of growth of the velocity mode.

The shadow image for the time t=29460 s ($t/\tau_{m}\simeq3.2\times10^{-2}$) for the $\tilde{R}=1.90\times10^{9}$ run is shown in Fig.4. In spite of the extremely large value of $\tilde{R}$, a stable spoke pattern is observed. We find this result rather striking. Indeed the $\tilde{R}$ value is roughly three orders of magnitude larger than the value $R\simeq10^{6}$ at which RBI single component systems exhibit a transition to turbulent convection and the associated lack of
spatial order. We will try below to give an explanation of this unexpected resilience against chaotic behavior. Some recent RBI numerical simulations for the mantle convection \cite{yuen2,yuen3} have shown that in the non-linear heat diffusion equation
\begin{equation}
\frac{\partial T}{\partial t}+\vec{u}\cdot\vec{\nabla}T=\kappa\nabla
^{2}T+\frac{\partial\kappa}{\partial T}(\nabla T)^{2}\label{heat}\end{equation}

a positive non-linear term $(\partial\kappa/\partial T)(\nabla T)^{2}$ implies
stabilization of the convective structures due to suppression of the small-scale boundary layer instabilities. This focusing effect is stronger where the regions of columnar flow exist. Conversely, a negative $\frac{\partial\kappa}{\partial T}$ favors the transition to turbulent
behavior. For the mantle, the positive sign contribution comes from the radiative component of the thermal diffusivity. The non-linear term $\frac{\partial\kappa}{\partial T}(\nabla T)^{2}$ in the temperature equation for RBI is replaced by $\frac{\partial D}{\partial c}(\nabla c)^{2}$ in the SDI concentration equation. In order to ascertain the sign of $\frac{\partial D}{\partial c}$ for the colloidal suspension we have performed Dynamic Light Scattering
measurements on the sample. We found that $\frac{1}{D}\frac{\partial D}{\partial c}$\ is $2.6$, somewhat larger than the Hard Sphere value of $0.67$ \cite{batchelor} as expected because of the electrostatic repulsion between the silica spheres. This suggests that a mechanism similar to that reported in the simulations of Refs.\cite{yuen2,yuen3} could be active in stabilizing the structures in Fig.1d against turbulence.

We are greatly indebted to D. S. Cannell and to D. A. Yuen for illuminating discussions. We thank R. Piazza and A. Guarino for the measurements of $k_{T}$ and $\frac{\partial D}{\partial c}$. D. Brogioli is thanked for help with the shadowgraph technique. Work partially supported by the Italian Space Agency (ASI) and by COFIN2001.

\pagebreak

\begin{figure*}
	\begin{center}
		\includegraphics{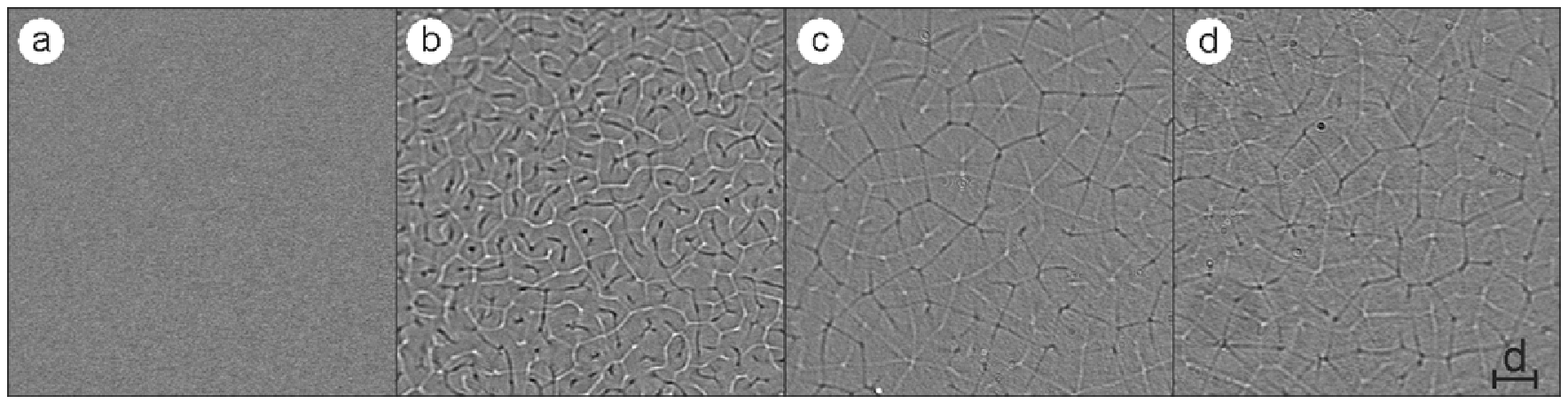}
	\end{center}
		\caption{Time evolution of shadowgraph images of SDI in a dilute colloidal suspension of 11 nm silica particles. The applied temperature difference is $13.22{{}^\circ}$C corresponding to $\tilde{R}=8.33\times10^{6}$. The images were taken (a) before turning on the servo at t=0, (b) at t=390 s, (c) at t=7200 s, and (d) at t=760000. The bar indicates the cell thickness d=0.98mm.}
	\label{sequence}
\end{figure*}

\begin{figure}
	\begin{center}
		\includegraphics{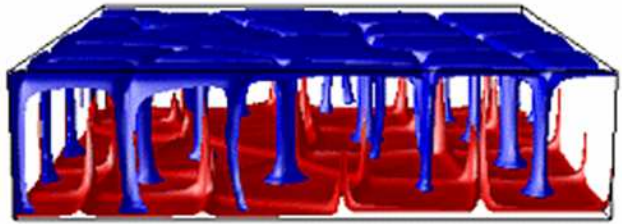}
	\end{center}
	\caption{Hot and cold isotherm surfaces from RBI simulation of Ref.\cite{yuen}, $R\simeq10^{6}$}
	\label{yuenfig}
\end{figure}

\begin{figure}
	\begin{center}
		\includegraphics{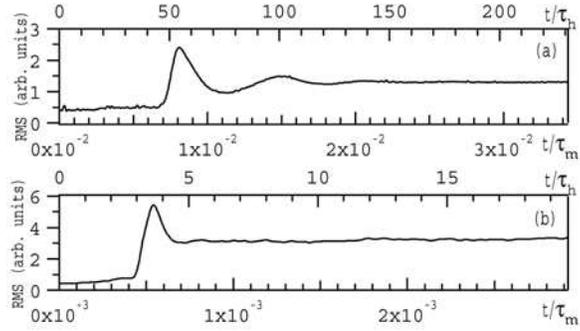}
	\end{center}
\caption{Time evolution of the RMS amplitude of the shadowgraph images during SDI in (a) 0.98 mm sample with $\Delta$T=13.22 K and $\tilde{R}=8.33\times10^{6}$ and (b) 4.5 mm thick sample, with $\Delta$T=31.13 K and $\tilde{R}\simeq1.9\times10^{9}$. The bottom (top) time axis has been rescaled with the mass (heat) diffusion time constant $t/\tau_{m}$ ($t/\tau_{h}$)}
	\label{rms}
\end{figure}

\begin{figure}
	\begin{center}
		\includegraphics{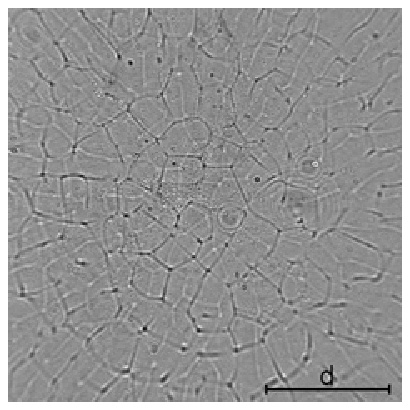}
	\end{center}
	\caption{Shadow image for the sample with thickness d=4.5mm, $\Delta$T=31.13 K and $\tilde{R}\simeq1.9\times10^{9}$ taken at $t/\tau_{m}\simeq3.2\times10^{-2}$. The bar indicates the cell thickness d=4.5mm.}
	\label{twothick}
\end{figure}


\begin{thebibliography}{9}
\bibitem {soret}A temperature gradient induces mass flow, so that at steady
state a stable concentration gradient $\vec{\nabla}c=-\frac{k_{T}}{T}%
\vec{\nabla}T$ is generated, where $k_{T}$ (which may have either sign) is the
thermal diffusion ratio. In this way $\psi=\frac{\beta}{\alpha}\frac{k_{T}}%
{T}$.
\end{thebibliography}
\end{document}